\documentclass[apl,amsmath,amssymb,reprint]{revtex4-1}
\usepackage{graphicx}
\usepackage{bm}
\bibstyle{aipnum4-1}

\begin{document}
\preprint{APL}
\title{Surfactant mediated growth of Ti/Ni multilayers}
\author{M. Gupta}
\email {mgupta@csr.res.in}
\author {S. M. Amir} \author {A. Gupta}
\affiliation{UGC-DAE Consortium for Scientific Research,
University Campus, Khandwa Road, Indore-452 001,India}
\author{J. Stahn}
\affiliation{Laboratory for Neutron Scattering,
Paul Scherrer Institut, CH-5232 Villigen PSI, Switzerland}

\date{\today}
\begin{abstract}
The surfactant mediated growth of Ti/Ni multilayers is studied.
They were prepared using ion beam sputtering at different adatom
energies. It was found that the interface roughness decreased
significantly when the multilayers were sputtered with Ag as
surfactant at an ion energy of 0.75\,keV. On the other hand, when
the ion energy was increased to 1\,keV, it resulted in enhanced
intermixing at the interfaces and no appreciable effect of Ag
surfactant could be observed. On the basis of the obtained results,
the influence of adatom energy on the surfactant mediated growth
mechanism is discussed.

\end{abstract}

\maketitle

Ti/Ni multilayers have found
applications in many emerging areas including neutron
mirrors,\cite{ju_APL07,padiyath_APL06,kumar_JAP1998} soft x-ray
optics,\cite{Mertins_AO98,Chaudhuri_TSF94,C_add_NIMA01} and shape
memory alloys for making microelectronic mechanical systems (MEMS)
devices.\cite{su_APL98} Also, this system is considered as a
model system for demonstrating the solid state amorphisation (SSA)
reaction.~\cite{SSA_PRB90,RG_TSF06,Basu_PRB09} Therefore the Ti/Ni
multilayer system is interesting both, from the basic as well as from the
application point of view. In particularly, the large difference
between the neutron scattering length densities (SLD) of Ti
($-1.95\cdot 10^{-6}\,\mathrm{\AA}^{-2}$) and
Ni ($9.41\cdot 10^{-6}\,\mathrm{\AA}^{-2}$) makes Ti/Ni multilayers
an ideal candidate in neutron optics. Attempts have been made to
further optimise this multilayer system either by increasing the
contrast by adding hydrogen \cite{ju_APL07,H_add_JAC92} or by
adding interdiffusion barriers such as Cr and
C \cite{C_add_NIMA01,C_add_NIMA04,Cr_add_NIMA06}.

The surface free energy ($\gamma$) for the average face of Ni, Ti
and Ag are $\gamma_{\mathrm{Ni}} = 2.4\,\mathrm{Jm^{-2}}$,
$\gamma_{\mathrm{Ti}} = 2.1\,\mathrm{Jm^{-2}}$ and
$\gamma_{\mathrm{Ag}} = 1.2\,\mathrm{Jm^{-2}}$,
respectively.~\cite{Tyson_SS77,Foiles_PRB86} A difference in the
$\gamma$ is known to produce multilayers with asymmetric interface
as wetting of the material with a lower $\gamma$ will take place
on a material with higher $\gamma$. In the opposite case
agglomeration due to de-wetting takes place. Therefore, from the
difference in the $\gamma$, it is expected that the Ti-on-Ni
interface should be smoother than the Ni-on-Ti interface in a
Ti/Ni multilayer. This asymmetry may result in increased roughness
and stress as the growth of the multilayer takes place. Such an
increase in roughness and stress leads to a decrease in the peak
reflectivity, and to an enhanced interdiffusion. In a neutron
monochromator or supermirror several hundreds of repetitions are
required for the optimum performance of the device. In this
situation the stress can buildup so intensely that the deposited
film may peel-off from the substrate as observed
experimentally.\cite{Senthil_PHYS.B97}

The addition of a surface active species (a {\em surfactant})
might balance the asymmetry of the interfaces.\cite{Cope_PRL89} A
surfactant essentially has a significantly lower $\gamma$ than the
multilayer components so it leads to a wetting and thus to
smoother interfaces. If in addition it is not incorporated in
either of the compounds, it will flow on top during the deposition
and thus allow for a layer-by-layer type growth throughout the
complete film.\cite{Barabasi_PRL1993} In the case of Ti/Ni
multilayers, no attempt has been made yet to study the surfactant
mediated growth, although surfactants have been used in other
types of multilayers e.g. Sb as surfactant in Si/Ge
multilayers,\cite{SiGe_APL05} O and Ag as surfactants in Cu/Co
multilayers,\cite{CoCu_PRL98,MG_CuCo}, etc. In addition, the role
of adatom energy on the surfactant mechanism has not been studied.


In this letter, we present the results of surfactant mediated
growth of Ti/Ni multilayers, prepared using ion beam sputtering
(IBS) at incident ion energy $E_\mathrm{Ar^+} =
0.75\,\mathrm{keV}$ and $1.00\,\mathrm{keV}$. Two otherwise
identical sets of samples have been prepared on silicon (100)
substrates at room temperature with following deposition sequences:\\[1ex]
 \begin{tabular}{rl}
  A:&$\mathrm{Ti(10\,nm)\,/\,[\,Ti(6\,nm)\,/\,Ni(4\,nm)]_{10}}$, \\
  B:&$\mathrm{Ti(10\,nm)\,/\, Ag(0.2\,nm)\,/\,[\,Ti(6\,nm)\,/\, Ni(4\,nm)\,]_{10}}$, \\
  C:&$\mathrm{Ti(10\,nm)\,/\, [\,Ti(6\,nm)\,/\,Ag(0.2\,nm)\,/\, Ni(4\,nm)\,]_{10}}$. \\
 \end{tabular}\\[1ex]

The ion beam produced with a RF-ion source (Veeco) was kept
incident at angle of 45\,$^{\circ}$ with respect to the target.
The base pressure was less than $5\cdot10^{-8}$\,mbar prior to the
deposition while the pressure during the deposition was $5\cdot
10^{-4}$\,mbar due to the flow of the Ar gas (purity 99.9995\%) in
the ion source and RF-neutralizer. Each target was pre-sputtered
for about 15\,minutes to remove contaminations from the target.
The pre-sputtering of Ti also helped to minimize the oxygen
partial pressure. The targets mounted on a rotary motion were
alternatively rotated to deposit the multilayer structures.

\begin{figure}  \center \vspace{-5mm}
\unitlength 1mm
\includegraphics [width=65mm,height=65mm]{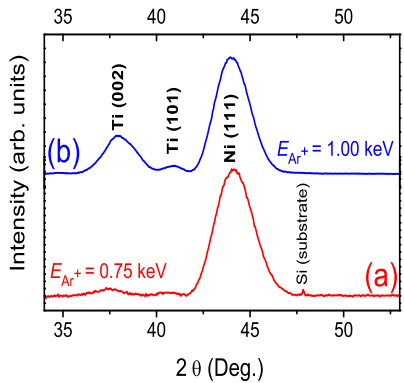}
\vspace{-0.4cm} \caption{\label{fig:fig1} (Color online) X-ray
(Cu-K$\alpha$) diffraction pattern of samples prepared with
$E_\mathrm{Ar^+} = 0.75\,\mathrm{keV}$ (a), and with
$E_\mathrm{Ar^+} = 1.00\,\mathrm{keV}$ (b), respectively.}
\end{figure}

The thickness of the deposited thin films was calibrated using
x-ray (Cu-K$\alpha$) reflectivity (XRR). X-ray diffraction
measurements were carried out to measure the structure of the
deposited samples. Fig.~\ref{fig:fig1} shows the XRD pattern of
the samples A. The peaks observed correspond to Ni (111), Ti (002)
and Ti (101). As can be seen, the peaks are relatively broader for
the sample prepared using a lower Ar$^+$ ion energy. The average
grain size can be calculated using the Scherrer formula: $t =
0.9\lambda/\Delta\theta \cos \theta$, where $t$ is the grain size,
$\Delta\theta$ is an angular width of the reflection, $\theta$ is
the Bragg angle, and $\lambda$ is the wavelength of the radiation
used. The grain size of the sample prepared using $E_\mathrm{Ar^+}
= 1.00\,\mathrm{keV}$ was 4.8$\pm$0.1\,nm which is somewhat larger
as compared to 4.2$\pm$0.1\,nm for the sample with
$E_\mathrm{Ar^+} = 0.75\,\mathrm{keV}$. All XRD patterns of
samples with varying surfactant contents but same incident ion
energy were identical.

\begin{figure}
\center \vspace{-8mm}
\includegraphics [width=85mm,height=70mm] {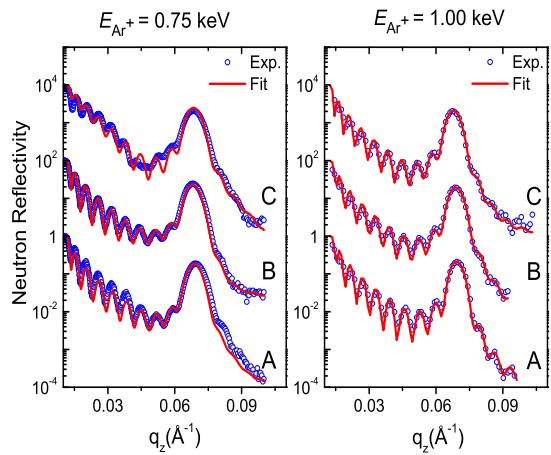}
\caption{\label{fig:fig2} (Color online) Neutron reflectivitiesy
of the Ti/Ni multilayers prepared with $E_\mathrm{Ar^+} =
0.75\,\mathrm{keV}$ (left), and with $E_\mathrm{Ar^+} =
1.00\,\mathrm{keV}$ (right). The reflectivities are scaled for
clarity.} \vspace{-4mm}
\end{figure}

\begin{figure}  \center
\includegraphics [width=85mm,height=85mm]{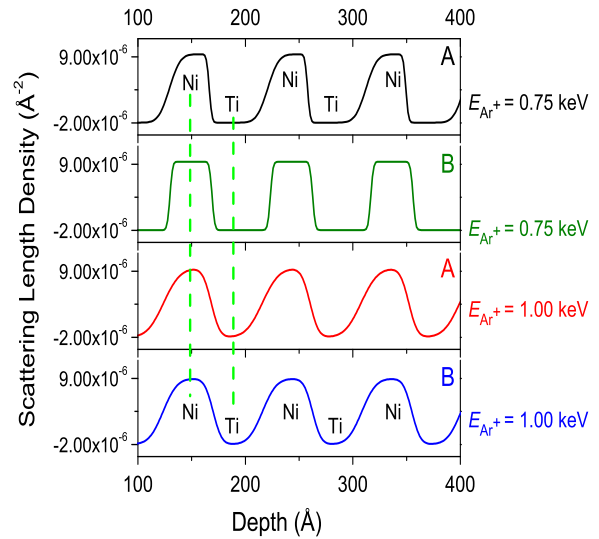}
\vspace{-4mm} \caption{\label{fig:fig3} (Color online) SLD depth
profiles for Ti/Ni multilayers prepared without surfactant (A),
and with Ag on the Ti buffer layer, only (B), for both sputter
energies $E_\mathrm{Ar^+} = 0.75\,\mathrm{keV}$ (top), and
$E_\mathrm{Ar^+} = 1.00\,\mathrm{keV}$ (bottom), respectively.}
\end{figure}

Neutron reflectivity (NR) measurements were carried out to
measure the depth profile of the SLDs.
The samples prepared with $E_\mathrm{Ar^+} = 0.75\,\mathrm{keV}$
were measured at the reflectometer AMOR at SINQ/PSI, Switzerland in
the time of flight mode \cite{Gupta_PJP04}, and
the ones with $E_\mathrm{Ar^+} = 1.00\,\mathrm{keV}$ in the monochromatic mode
($\lambda = 4.4\,\mathrm{\AA}$) at Superadam
at ILL, France.
The resulting curves are shown in Fig.\,\ref{fig:fig2}.
They were fitted using Parratt's
formulism.\cite{Parratt32}
While the NR pattern for samples A and B with
$E_\mathrm{Ar^+} = 0.75\,\mathrm{keV}$
can be fitted well assuming their nominal
structures, the one for sample C shows a
broad hump riding on top of total thickness oscillations. This
hump could be fitted assuming a layer of about 2\,nm on
top of the multilayer structure with a SLD close to that of bulk Ag.
It looks like all the Ag from all Ti-Ni interfaces accumulated
on top of the film.
The details of fitting are given in table~\ref{tab1}. As can be seen
there, the reflected intensity ($\mathcal{R}$) at the Bragg
peak ($q_z = 0.068\,\mathrm{\AA}^{-1}$) for the sample prepared at
0.75\,keV without any surfactant was 18.6\% while it increases to
24.3\% when Ag was added at the buffer Ti layer, and
23.7\% when the surfactant was introduced at each Ti layer.
Whereas, in case of the 1\,keV sample $\mathcal{R}$ only
increases marginally with addition of Ag.

The SLD depth profiles for samples A and B, obtained by fitting of
the NR data, are plotted in fig.~\ref{fig:fig3}. The SLD depth
profile for the reference sample A deposited at $E_\mathrm{Ar^+} =
0.75\,\mathrm{keV}$, shows an asymmetry with an abrupt Ti-on-Ni
and a skewed Ni-on-Ti interface. As expected, Ti with lower
$\gamma$ should wet the surface of Ni, while Ni will agglomerate
on Ti resulting in a rougher Ni-on-Ti interface. The obtained
interface rms roughnesses ($\sigma$) are given in table
\ref{tab1}. A decrease in the interface roughness results in an
increased $\mathcal{R}$. By adding Ag either on the buffer Ti
layer or at each Ti layer, the SLD profile becomes symmetric and
the overall roughness decreases. In the case of samples prepared
at $E_\mathrm{Ar^+} = 1.00\,\mathrm{keV}$ the results are markedly
different as $\mathcal{R}$ only increases marginally when the
surfactant was used. The SLD depth profiles are also broader with
appreciably higher roughnesses. Still, the roughness of Ni-on-Ti
interface of $\approx 1\,\mathrm{nm}$ is higher as compared to
$\approx 0.6\,\mathrm{nm}$ for Ti-on-Ni interface. Symmetric and
smooth interfaces obtained using Ag surfactant at $E_\mathrm{Ar^+}
= 0.75\,\mathrm{keV}$ are indicative of layer-by-layer type
growth~\cite{Barabasi_PRL1993} whereas the growth of Ti/Ni
multilayers at $E_\mathrm{Ar^+} = 1.00\,\mathrm{keV}$ seems to
remain unaffected with addition of Ag surfactant.


\begin{table}
\caption{Parameters obtained by fitting of neutron reflectivity
pattern of Ti/Ni multilayers.} \label{tab1}
\begin{tabular}{c|ccc|ccc}
\toprule
Ar$^+$ ion energy &&0.75 keV &&& 1 keV &\\
\colrule
Samples&A& B& C& A& B& C\\
\colrule
$\mathcal{R}$ (\%)&18.6& 24.4& 23.7& 16.3& 17.4 & 18.1\\
\colrule
$\sigma_{\mathrm{[Ni-on-Ti]}}$ (nm) &0.8 &0.2 &0.2 &  1.20&  1.03&  0.90 \\
\colrule
$\sigma_{\mathrm{[Ti-on-Ni]}}$ (nm) &0.2 &0.2 & 0.2& 0.64&  0.64&  0.60 \\
\botrule
\end{tabular}
\end{table}

The observed results can be understood in terms of energy of the
sputtered atoms (or adatoms, $E_\mathrm{ad}$). Although
$E_\mathrm{ad}$ is difficult to calculate precisely, the mean
values can be estimated using SRIM calculations.~\cite{TRIM,SRIM}
The calculated values of $E_\mathrm{ad}$ (eV/atom) at
$E_\mathrm{Ar^+} = 0.75\,\mathrm{keV}$ for Ni, Ti and Ag are 22.6,
25.3 and 20.7, respectively whereas at $E_\mathrm{Ar^+} =
1.00\,\mathrm{keV}$ they increase to 23, 33.3 and 22.2 for Ni, Ti
and Ag, respectively. As can be seen while the $E_\mathrm{ad}$ for
Ni and Ag increases marginally with an increase in incident ion
energy, for Ti it increases appreciably. This may results in an
increased mobility of the Ti atoms during condensation at the
substrate. As supported by the XRD measurements, a larger grain
size and sharper Ti peaks for the sample prepared at higher Ar$^+$
ion energy may be due to the increased mobility of the Ti adatoms.
This situation may lead to an enhanced intermixing at the
interfaces and the surfactant atoms may migrate in the host Ti
layer and therefore will not be able to float-off to the surface.
Therefore, the addition of surfactant atoms will have no or only
marginal effect on interface asymmetry as observed in this case.
This hypotheses is supported by the absence of the top Ag layer
for sample C in this case.

In conclusion, we observe smooth and symmetric interfaces,
indicative of layer-by-layer type growth, in Ti/Ni multilayers
prepared using Ag as surfactant at a deposition energy
$E_\mathrm{Ar^+} = 0.75\,\mathrm{keV}$. However, when this energy
is increased at $E_\mathrm{Ar^+} = 1.00\,\mathrm{keV}$ there is
only a marginal effect due to the increased mobility of Ti atoms.
Therefore, the energy of the adatoms plays a significant role in
the mechanism of the surfactant mediated growth of Ti/Ni
multilayers.

We acknowledge DST, Government of India for providing financial
support to carry out NR experiments under its scheme `Utilisation
of International Synchrotron Radiation and Neutron Scattering
facilities'. A part of this work was performed under the
Indo-Swiss Joint Research Programme with grant no.
INT/SWISS/JUAF(9)/2009. Thanks to A.\,Wildes for the help provided
in the NR measurements at ILL and to S.\,Potdar in sample
preparation.


%

\end{document}